\documentclass{ws-p8-50x6-00}

\begin{document}

\title{Soft QCD Modeling of Meson Electromagnetic Form Factors}

\author{Peter C. Tandy}

\address{Center for Nuclear Research, Department of Physics, Kent State 
University, Kent, OH 44242, USA\\ 
E-mail: tandy@cnr2.kent.edu}


\maketitle

\abstracts{We summarize recent progress in soft QCD modeling based on the 
set of Dyson--Schwinger equations truncated to ladder-rainbow level.   We 
pay particular attention to electromagnetic elastic and transition form 
factors of the pion.   This covariant approach accommodates quark confinement 
and implements the QCD one-loop renormalization group behavior.  The dressed 
quark propagators are compared to the most recent lattice-QCD data.}

\section{Introduction}

High energy electroweak probes utilize perturbative QCD  to quantify the
structure of hadrons in terms of  parton
structure functions.  Their detailed behavior samples various aspects
the quark-gluon dynamics of the hadron bound state.  This information is 
intrinsically non-perturbative and the unraveling of it from sets of structure 
functions will require a connection be made with non-perturbative QCD 
calculations and models.   Although some lattice QCD studies have begun to 
produce moments of  structure functions~\cite{Gockeler:1996wg}, the 
opportunities are very limited at present.  
The most extensive  hadronic models were designed to study the mass spectrum 
and decays and usually contain
elements that limit their use for high energy lepton scattering.  Examples
include:  non-relativistic kinematics, a lack of manifest Lorentz covariance, 
no quark sea, no dynamical gluons, no renormalization group behavior for change
of scale, and no confinement of quarks. 
Studies of deep inelastic scattering  within the
Nambu--Jona-Lasinio model~\cite{Weigel:1999pc} have helped clarify some of the 
issues confronting work within a quark field theory format.   

Here we summarize recent progress in soft QCD modeling based on the 
set of Dyson--Schwinger equations [DSEs] of the 
theory~\cite{Roberts:2000aa,Alkofer:2000wg}; we pay particular
attention to electromagnetic couplings to the pion.   This covariant approach
accommodates quark confinement and implements the QCD one-loop renormalization 
group behavior.   Deep inelastic scattering phenomena have not been treated
with the present model; a simplified related approach that retains all the 
essential features has recently produced excellent results for the pion 
valence quark distribution amplitude~\cite{Hecht:2001xa}.

Meson electromagnetic form factors in impulse approximation are
described by two diagrams, in one the photon couples to the quark, in
the other to the antiquark.  A form factor can be associated with each 
diagram.  For example, the photon coupling to the antiquark of a flavor
$a\bar{b}$ pseudoscalar meson produces~\cite{MTpiK00}
\begin{eqnarray}
 2\,P_\nu\,F_{a\bar{b}\bar{b}}(Q^2) &=&
        N_c\int^\Lambda\!\!\frac{d^4q}{(2\pi)^4} \,{\rm Tr}\big[ 
               S^a(q) \, \Gamma_{\rm ps}^{a\bar{b}}(q,q_+;P_-) \, S^b(q_+)
\nonumber \\ &&{}\times
         i \Gamma^{b}_\nu(q_+,q_-)\, S^b(q_-) \,
        \bar\Gamma_{\rm ps}^{a\bar{b}}(q_-,q;-P_+) \big] \;, 
\label{impulse}
\end{eqnarray}
where \mbox{$q = k+\frac{1}{2}P$}, 
\mbox{$q_\pm = k-\frac{1}{2}P \pm \frac{1}{2}Q$},  
\mbox{$P_\pm = P \pm \frac{1}{2}Q$}.  We work in the isospin symmetry
limit, so for the pion  we have \mbox{$F_{\pi}(Q^2) =
F_{u\bar{u}u}(Q^2)$}.  The charged and neutral kaon form factors are 
given by \mbox{$F_{K^+}= \frac{2}{3}F_{u\bar{s}u} + 
\frac{1}{3}F_{u\bar s \bar s}$} and \mbox{$F_{K^0} = 
-\frac{1}{3}F_{d\bar{s}d} + \frac{1}{3}F_{d\bar s\bar s}$} respectively.
To implement such a study, one needs a consistent QCD modeling of the 
photon-quark dressed vertex $\Gamma_\nu$, the dressed quark propagator $S(q)$, 
and the meson Bethe-Salpeter (BS) amplitude $\Gamma_{ps}$.  The first two 
are related by the vector Ward-Takahashi identity[WTI]; the second two are 
related by the axial-vector WTI. 

The DSE for the renormalized quark propagator in Euclidean space is
\begin{equation}
\label{gendse}
 S(p)^{-1} = i \, Z_2\, /\!\!\!p + Z_4\,m(\mu) + 
        Z_1 \int^\Lambda\!\!\!\frac{d^4q}{(2\pi)^4} \,g^2 D_{\mu\nu}(k) 
        \frac{\lambda^a}{2}\gamma_\mu S(q)\Gamma^a_\nu(q,p) \,,
\end{equation}
where $D_{\mu\nu}(k)$ is the dressed-gluon propagator,
$\Gamma^a_\nu(q;p)$ is the dressed-quark-gluon vertex, and \mbox{$k=p-q$}.  
The  solution  has the form
\mbox{$S(p)^{-1} = i /\!\!\! p A(p^2) + B(p^2)$} and is renormalized 
at spacelike $\mu^2$ according to \mbox{$A(\mu^2)=1$}, and
\mbox{$B(\mu^2)=m(\mu)$}, where $m(\mu)$ is the current quark mass.
The notation $\int^{\Lambda}$ represents a translationally invariant
regularization at scale $\Lambda$.  The same regularization occurs 
consistently at all stages as indicated in Eq.~(\ref{impulse}).
One takes \mbox{$\Lambda \to \infty$} as the final step.  

The $q\bar q\gamma$ vertex \mbox{$\Gamma_\mu(p_+,p_-)$} corresponding to total
momentum \mbox{$Q = p_+ - p_-$} satisfies the Bethe-Salpeter equation [BSE]
\begin{equation}
 \Gamma_\mu(p_+,p_-) = Z_2 \, \gamma_\mu + 
        \int^\Lambda\!\!\!\frac{d^4q}{(2\pi)^4} \, K(p,q;Q) 
        \;S(q_+) \, \Gamma_\mu(q_+,q_-) \, S(q_-)\, ,
\label{verBSE}
\end{equation}
where we use the notation $p_+ = p + \eta Q$ and $p_- = p - (1-\eta) Q$ for 
the outgoing and incoming quark momenta respectively at the vertex.   Here 
$\eta$, which specifies how the total momentum is shared between quark and 
antiquark (thus defining a relative momentum $p$), is arbitrary and physical 
observables should not depend on it.  The kernel $K$ is the renormalized, 
amputated $q\bar q$ scattering kernel that is irreducible with respect to 
a pair of $q\bar q$ lines.   Other electroweak processes~\cite{Ji:2001pj} 
require, e.g., the dressed  $q\bar q W$ vertex which is defined in an 
analogous way.  

When the homogeneous version of Eq.~(\ref{verBSE}) has vector solutions at 
(discrete) timelike  total momenta \mbox{$Q^2=-m^2$}, these are the vector
meson bound states.   The vector vertex $\Gamma_\mu(p_+,p_-)$ will have 
simple poles at those locations.   This correspondance holds for all other
transformation characters (pseudoscalar, axial vector, etc), labelled by 
quantum numbers \mbox{$J^{PC}$}.   The homogeneous solutions 
$\Gamma_M(p_+,p_-)$ are the bound state BS amplitudes 
and they are normalized in the  canonical way.
Explicit representations of the $\Gamma_M(p_+,p_-)$ require an expansion
in a complete set of covariants constructed from gamma matrices and momenta.
For example,the general representation for pseudoscalar bound states 
is~\cite{Maris:1997tm}
\begin{equation}
\label{genpion}
\Gamma_{ps}(k_+,k_-;P) = \gamma_5 \big[ i E +  \;/\!\!\!\! P \, F 
  + \,/\!\!\!k \, G + \sigma_{\mu\nu}\,k_\mu P_\nu \,H \big]\,,
\end{equation}
where the invariant amplitudes $E$, $F$, $G$ and $H$ are Lorentz
scalar functions $f(k^2;k\cdot P;\eta)$.  Note that these functions
depend on the momentum partitioning parameter $\eta$ because  the total and 
relative momenta have been employed in the covariants. However, physical 
observables are 
independent of this parameter; this is verified numerically within the
present approach as long as the set of employed covariants is 
complete~\cite{MTpiK00}.
%
\section{Ladder-Rainbow Model}
To solve the BSE, we use a ladder truncation, with an effective
quark-antiquark interaction that reduces to the perturbative running
coupling at large momenta~\cite{Maris:1997tm,Maris:1999nt}.  In
conjunction with the rainbow truncation for the quark DSE, the ladder
truncation of the BSE preserves both the vector WTI for
the $q\bar q\gamma$ vertex and the axial-vector WTI.  The latter ensures
the existence of almost massless pseudoscalar mesons which are the Goldstone 
bosons connected with dynamical
chiral symmetry breaking~\cite{Maris:1997tm}.  In combination with the
impulse approximation, this ladder-rainbow truncation ensures
electromagnetic current conservation~\cite{MTpiK00}.

The ladder truncation of the BSE, Eq.~(\ref{verBSE}), is
\mbox{$K(p,q;P) \to $} \mbox{$-{\cal G}(k^2)\, D_{\mu\nu}^{\rm free}(k)$}
\mbox{$\textstyle{\frac{\lambda^i}{2}}\gamma_\mu \otimes
        \textstyle{\frac{\lambda^i}{2}}\gamma_\nu$}
where \mbox{$k =p-q$}, and \mbox{$D_{\mu\nu}^{\rm free}$} is the free gluon 
propagator in
Landau gauge.  The corresponding rainbow truncation of the quark DSE,
Eq.~(\ref{gendse}), is given by \mbox{$\Gamma^i_\nu(q,p) \rightarrow 
\gamma_\nu\lambda^i/2$} together with \mbox{$g^2 D_{\mu \nu}(k) 
\rightarrow {\cal G}(k^2) D_{\mu\nu}^{\rm free}(k) $}.
This truncation was found to be particularly suitable for the flavor
octet pseudoscalar and vector mesons since the next-order
contributions in a quark-gluon skeleton graph expansion, have a
significant amount of cancellation between repulsive and attractive
corrections~\cite{Bender:1996bb}.  
\begin{table}[ht]
\caption{Results~\protect\cite{MTpiK00,Maris:1999nt,Maris:2000bh,MTinprogress,Maris:2000wz} for 
light mesons in GeV compared to 
experiment~\protect\cite{PDG}.  \label{tab:masses} }
\vspace{0.1cm}
\begin{center}
\begin{tabular}{l|cll}
        & \multicolumn{2}{l}{expt. (estimates)} 
        &  calc. (\underline{ fitted}) \\ \hline
$m^{u=d}_{\mu=1 {\rm GeV}}$       
                && 5 - 10 MeV  &  5.5 MeV    \\
$m^{s}_{\mu=1 {\rm GeV}}$                        
                &&100 - 300 MeV&  125 MeV     \\ \hline
- $\langle \bar q q \rangle^0_{\mu}$
                && (0.236 GeV)$^3$ & (\underline{ 0.241} GeV)$^3$ \\
$m_\pi$         &&  0.1385 GeV &   \underline{ 0.138} \\
$f_\pi$         &&  0.0924 GeV &   \underline{ 0.093} \\
$m_K$           &&  0.496 GeV  &   \underline{ 0.497} \\
$f_K$           &&  0.113 GeV  &   0.109        \\ \hline
$m_\rho$        &&  0.770 GeV  &   0.742        \\
$f_\rho$        &&  0.216 GeV  &   0.207        \\
$m_{K^\star}$   &&  0.892 GeV  &   0.936        \\
$f_{K^\star}$   &&  0.225 GeV  &   0.241        \\
$m_\phi$        &&  1.020 GeV  &   1.072        \\
$f_\phi$        &&  0.236 GeV  &   0.259        \\ \hline
$r^2_\pi$       &&  0.44 fm$^2$ &  0.45         \\
$r^2_{K^+}$     &&  0.34 fm$^2$ &  0.38         \\
$r^2_{K^0}$     && -0.054 fm$^2$& -0.086        \\
$r^2_{\pi\gamma\gamma}$&& 0.42 fm$^2$&  0.39    \\ \hline
$g_{\pi\gamma\gamma}$&&  0.50   &  0.50         \\
$g_{\rho\pi\gamma}$&&  0.57     &  0.54         \\
$g_{\rho\pi\pi}$&&  6.02       &   4.85         \\
$g_{\phi K K}$  &&  4.64       &   4.63         \\
$g_{K^{\star +} \pi^+ K^0}$ &&  4.53  & 4.6   \\\hline
\end{tabular}
\end{center}
\end{table}

The model is completely specified once a form is chosen for the
``effective coupling'' ${\cal G}(k^2)$.  We employ the
Ansatz~\cite{Maris:1999nt}
\begin{eqnarray}
\label{gvk2}
\frac{{\cal G}(k^2)}{k^2} &=&
        \frac{4\pi^2\, D \,k^2}{\omega^6} \, {\rm e}^{-k^2/\omega^2}
        + \frac{ 4\pi^2\, \gamma_m \; {\cal F}(k^2)}
        {\textstyle{\frac{1}{2}} \ln\left[\tau + 
        \left(1 + k^2/\Lambda_{\rm QCD}^2\right)^2\right]} \;,
\end{eqnarray}
with \mbox{$\gamma_m=12/(33-2N_f)$} and
\mbox{${\cal F}(s)=(1 - \exp\frac{-s}{4 m_t^2})/s$}.  
The ultraviolet behavior is chosen to be that of the QCD running
coupling $\alpha(k^2)$; the ladder-rainbow truncation then generates
the correct perturbative QCD structure of the DSE-BSE system of
equations.  The first term implements the strong infrared
enhancement
in the region \mbox{$0 < k^2 < 1\,{\rm GeV}^2$} phenomenologically
required to produce a realistic value for the chiral condensate.   In 
recent years, there has been some progress in
understanding the interplay between ghosts and gluons in Landau-gauge
QCD, indicating that ghosts could play an important role for this
enhancement~\cite{Alkofer:2000wg}.  However, this
is not yet ready for incorporation into phenomenological studies of hadron
properties.   We use \mbox{$m_t=0.5\,{\rm GeV}$},
\mbox{$\tau={\rm e}^2-1$}, \mbox{$N_f=4$}, \mbox{$\Lambda_{\rm QCD} =
0.234\,{\rm GeV}$}, and a renormalization scale \mbox{$\mu=19\,{\rm
GeV}$} which is well into the perturbative
domain~\cite{Maris:1997tm,Maris:1999nt}.  The remaining parameters,
\mbox{$\omega = 0.4\,{\rm GeV}$} and
\mbox{$D=0.93\,{\rm GeV}^2$} along with the quark masses, are fitted to
give a good description of $\langle \bar{q} q \rangle$, $m_{\pi/K}$ and
$f_{\pi}$.   The value of $f_K$, the ground state vector masses and decay 
constants, and the other quantities in Table~\ref{tab:masses} are then produced
without any free parameters~\cite{MTpiK00,Maris:1999nt,Maris:2000bh}.
Note that the complete set of covariants of the BS amplitudes, see
Eq.~(\ref{genpion}), are needed to satisfy the axial-vector
WTI~\cite{Maris:1997tm}.   The pseudovector amplitudes
$F$ and $G$  contribute about 30\% of the masses and decay constants of the
pseudoscalars; they dominate the asymptotic behavior of the pion charge form
factor~\cite{Maris:1998hc}.  
%
\begin{figure}[ht]
\hspace*{-0.5cm}\psfig{figure=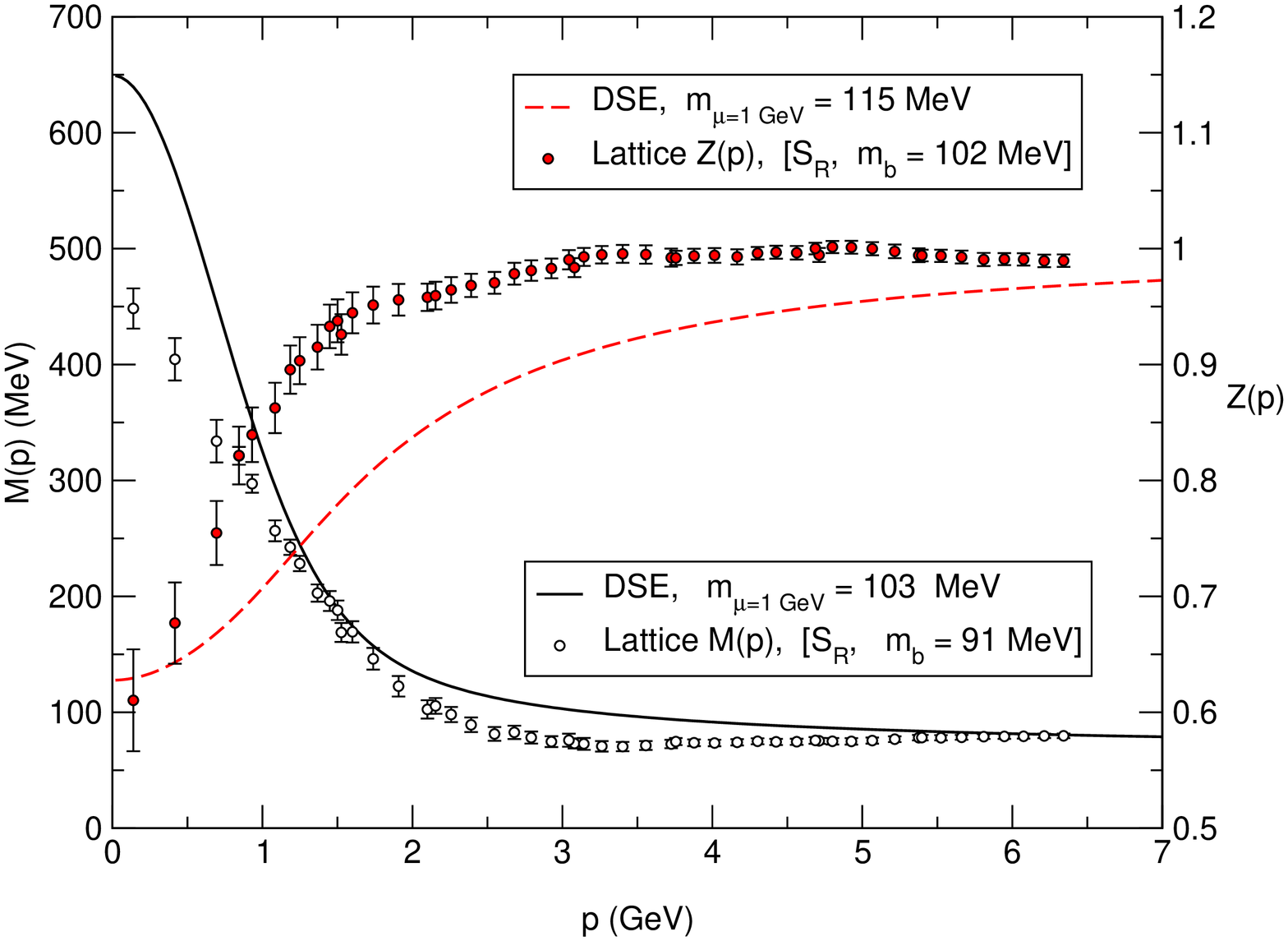,height=7.0cm,angle=-90}
\hspace*{-0.2cm}\psfig{figure=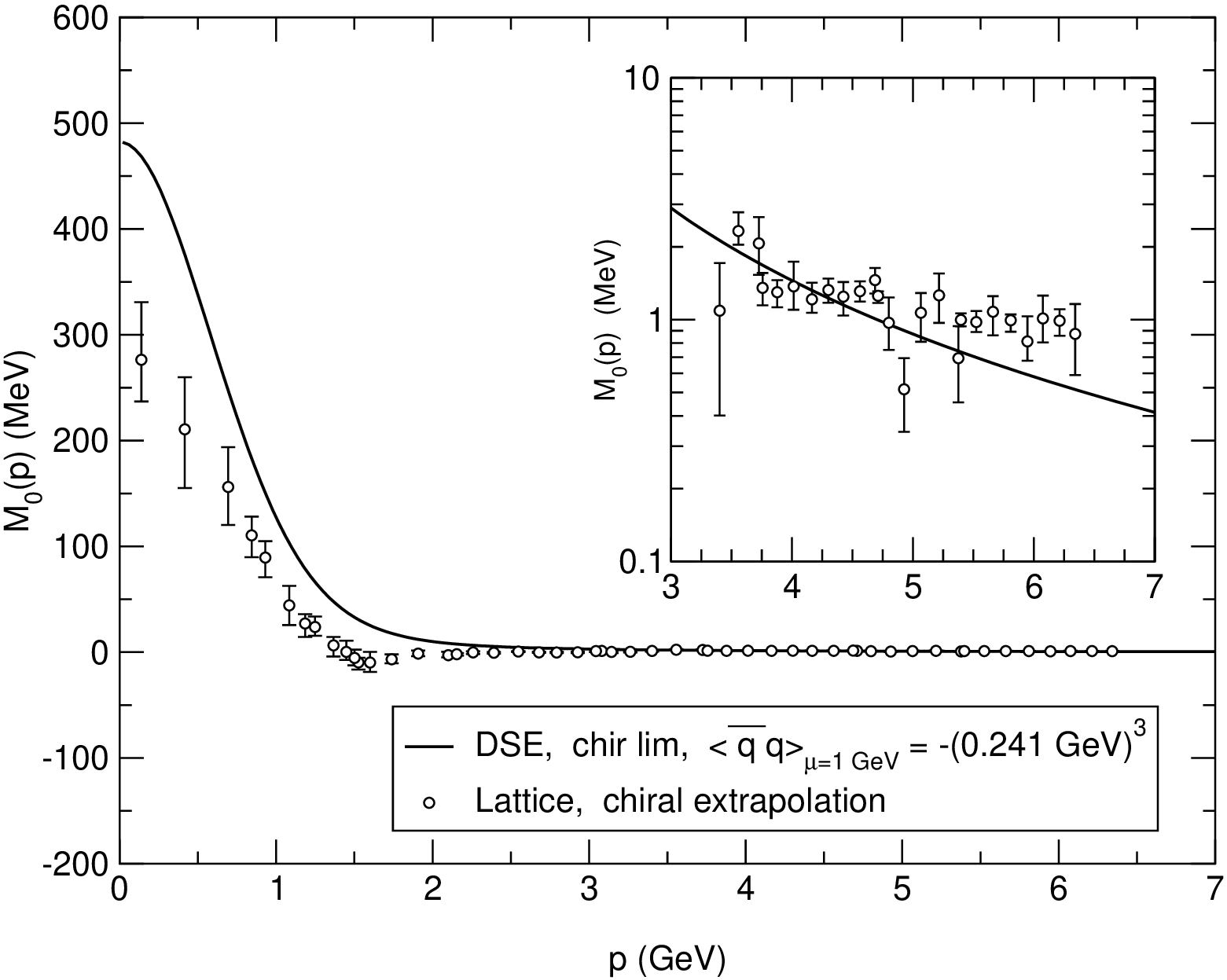,height=7.0cm,angle=-90}
\caption{{\it Left Panel}: DSE solution~\protect\cite{Maris:1999nt} for quark 
propagator amplitudes compared to recent lattice 
data~\protect\cite{Skullerud:2000un}. {\it Right Panel}: The chiral limit
DSE mass function compared to the lattice chiral
extrapolation~\protect\cite{Skullerud:2000un}. 
\label{fig:qrkprp}}
\end{figure} 

In Fig.~\ref{fig:qrkprp} we compare the DSE model~\cite{Maris:1999nt} 
propagator amplitudes
defined by \mbox{$S(p) =Z(p^2)[ i /\!\!\! p  + M(p^2)]^{-1}$} with the most
recent results from lattice QCD~\cite{Skullerud:2000un}.  In contrast
to a previous comparison~\cite{Maris:2000zf}, the data here has more lattice
artifacts removed  and the DSE calculations use the indicated current mass
values to match the lattice mass function at $6$~GeV.  There is agreement 
in the qualitative infrared structure of the mass function particularly in the 
way the infrared enhancement sets in.   Since the lattice
simulation produces the regulated but un-renormalized propagator, the scale
of the field renormalization function $Z$ is arbitrary and only the shape 
is a meaningful comparison.  For this reason, we have scaled the lattice data 
for $Z$ so that \mbox{$Z(5~{\rm GeV})=1$}.  The ladder-rainbow DSE model 
typically produces a $Z$
that saturates much slower than does the lattice $Z$; this may signal a
deficiency of the bare gluon-quark vertex.  The lattice 
work~\cite{Skullerud:2000un} also produced a linear extrapolation to the
chiral limit mass function $M_0(p)$ and we compare this to 
the DSE result with the insert emphasizing high momentum.
The known one-loop renormalization group UV behavior of chiral QCD, which 
is preserved by this DSE model, is
\begin{equation}
 M_0(p^2) \simeq \frac{2\pi^2\gamma_m}{3}\,
        \frac{-\,\langle \bar q q \rangle^0}{p^2 \left(
        \frac{1}{2}\ln\left[\frac{p^2}{\Lambda_{\rm QCD}^2}
                                        \right] \right)^{1-\gamma_m}}\,,
\end{equation}
with $\langle \bar q q \rangle^0$ being the renormalization-point-independent
chiral condensate~\cite{Maris:1997tm}.   Although the lattice data does not 
distinguish clearly between this fall-off and a flat behavior, the scale
of the data is consistent with a condensate that is less than 10\% higher
than that of the DSE model.   That is, we make a tentative assignment of
\mbox{$\langle \bar{q} q \rangle_{\mu=1~{\rm GeV}} = -(248~{\rm GeV})^3$}
for this lattice data.
\section{Meson Elastic and Transition Form Factors}
\begin{figure}[ht]
\hspace*{\fill}\psfig{figure=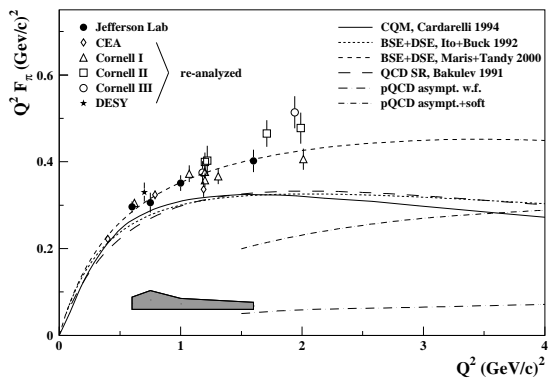,height=5.0cm}\hspace*{\fill}
\caption{Our results for the pion charge form 
factor~\protect\cite{MTpiK00} compared to new data from 
JLab~\protect\cite{Volmer} including a re-analysis of older data. Also shown 
are various earlier model results.  Figure taken from the PhD 
thesis of J. Volmer~\protect\cite{Volmer}.   \label{fig:Fpi_Jlab}}
\end{figure}

Our results~\cite{MTpiK00} for $F_\pi(Q^2)$  are shown
in Fig.~\ref{fig:Fpi_Jlab};  the charge radii are given in
Table~\ref{tab:masses}.  
%
Up to about $Q^2 = 3\,{\rm GeV}^2$, our results for $F_\pi(Q^2)$ and $F_K(Q^2)$
can be fitted quite well by a monopole~\cite{MTpiK00}.  Asymptotically, 
the behavior is \mbox{$Q^2 F(Q^2) \rightarrow c$} up to
logarithmic corrections.  However, numerical limitations prevent us
from accurately determining these constants. Around $Q^2 = 3\,{\rm
GeV}^2$, our result for $Q^2 F_\pi$ is well above the pQCD
result~\cite{Farrar:1979aw} \mbox{$16 \pi f_\pi^2 \alpha_s(Q^2) \sim
0.2 \,{\rm GeV}^2$}, and clearly not yet asymptotic.  The calculated 
time-like electromagnetic form factor displays vector meson bound state
poles from which one can extract the strong coupling constants 
$g_{\rho\pi\pi}$ and $g_{\phi K K}$ respectively~\cite{MTinprogress}.  The 
results are given in Table~\ref{tab:masses} and are close to the experimental 
data and to the results from direct evaluation~\cite{JMT01prep}.

\begin{figure}[ht]
\vspace*{0.2cm}

\psfig{figure=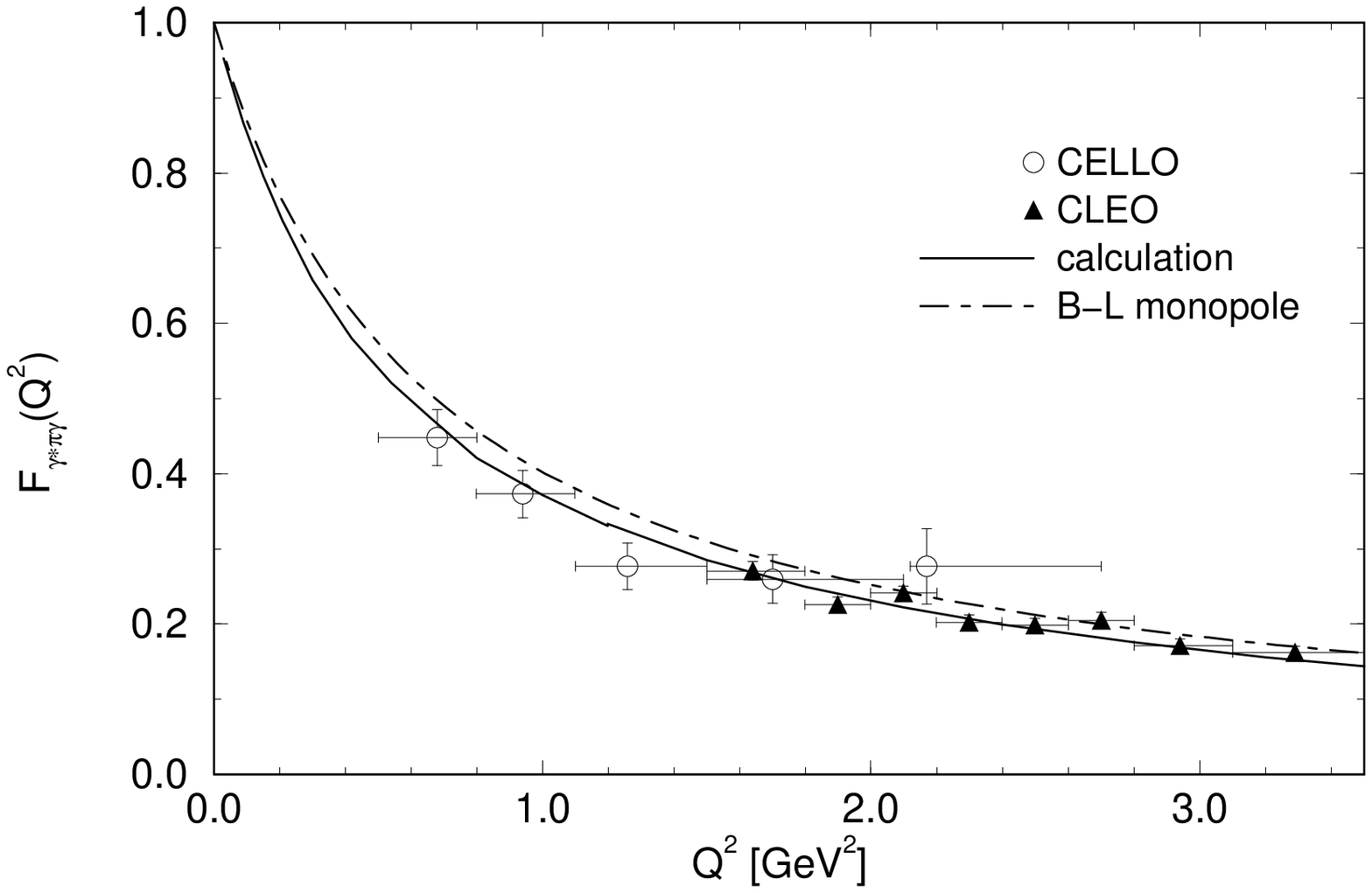,height=5.2cm,width=5.7cm}
\vspace*{-6.2cm}

\hspace*{6.1cm}\psfig{figure=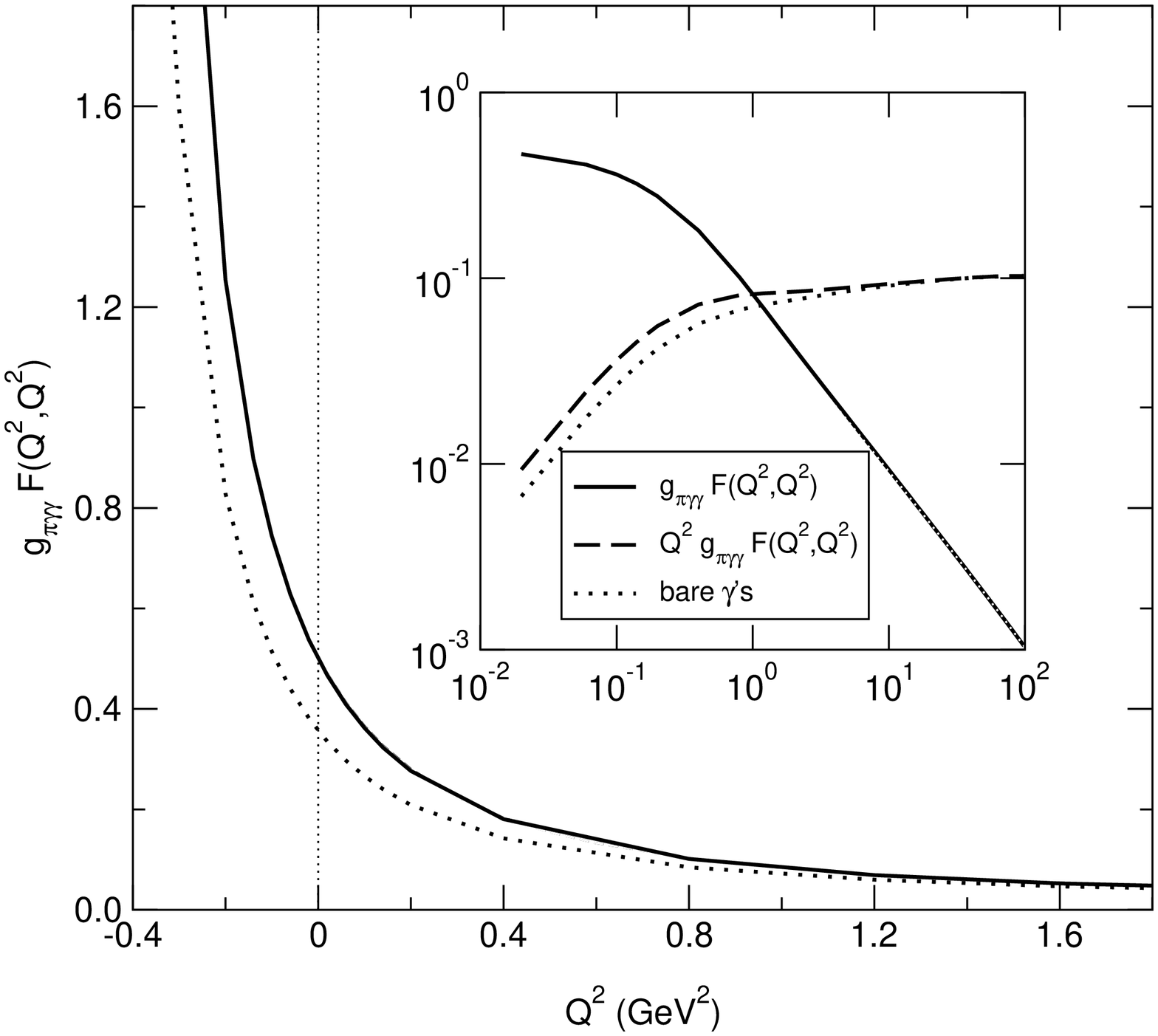,height=7.1cm,angle=-90}
\caption{{\it Left Panel}: The $\gamma^\star\,\pi\gamma$ form 
factor~\protect\cite{Maris:2000wz}, with data from CLEO and
CELLO~\protect\cite{cellocleo}. {\it Right Panel}: The form factor for 
$\gamma^\star\,\pi\gamma^\star$ for equal virtuality with the insert 
showing the asymptotic behavior of $Q^2 F(Q^2)$. \label{fig:piggff}}
\end{figure}

The impulse approximation for the $\gamma^\star\pi\gamma$ vertex 
with $\gamma^\star$ momentum $Q$ is
\begin{eqnarray}
\lefteqn{ \Lambda_{\mu\nu}(P,Q)=i\frac{\alpha }{\pi f_{\pi }}
        \,\epsilon_{\mu \nu \alpha \beta }\,P_{\alpha }Q_{\beta }
        \, g_{\pi\gamma\gamma}\,F_{\gamma^\star\pi\gamma}(Q^2)  } \\
& & \nonumber
        =\frac{N_c}{3}\, \int\!\frac{d^4q}{(2\pi)^4}
        {\rm Tr}\left[S(q)\, i\Gamma_\nu (q,q')\,S(q')\, 
                i\Gamma_\mu (q',q'')\,S(q'')\,\Gamma_\pi(q'',q;P)\right] \;.
\end{eqnarray}
where the momenta follow from momentum conservation.  In the chiral
limit, the value at $Q^2 = 0$, corresponding to the decay \mbox{$\pi^0
\rightarrow \gamma\gamma$}, is given by the axial anomaly and its value
\mbox{$g^{0}_{\pi\gamma\gamma}=\frac{1}{2}$} is a direct consequence of 
only gauge invariance and chiral symmetry; this value is reproduced by
our calculations~\cite{Maris:1998hc} and corresponds well with the
experimental width of $7.7~{\rm eV}$.  In Fig.~\ref{fig:piggff}~(left) we show
our results~\protect\cite{Maris:2000wz} with realistic quark masses, 
normalized to the experimental $g_{\pi\gamma\gamma}$.
For such moderate $Q^2$, both the data and our DSE model are close to
the monopole shape fitted to the Brodsky--Lepage  asymptotic
limit~\cite{Lepage:1980fj} \mbox{$F \sim 8 \pi^2 f_\pi^2$} from  pQCD in 
the factorized approximation.  As pointed out recently~\cite{Anikin:2000cx}, 
it is advantageous
to consider the process where both photons are virtual;  the light-cone
operator product expansion yields~\cite{Lepage:1980fj,Chase:1980ck}
\begin{equation}
g_{\pi\gamma\gamma} \,F_{\pi\gamma\gamma}(Q_1^2,Q_2^2) \to 
4\pi^2 f_\pi^2 \;\big\{ \frac{J(\omega)}{2(Q_1^2+Q_2^2)} \,+\,
   {\cal O}(\frac{\alpha_s}{\pi}, \frac{1}{Q^4}) \big\} \; .     
\label{lcope}
\end{equation}
Here the asymmetry is  
\mbox{$\omega = (Q_1^2 - Q_2^2)/(Q_1^2 + Q_2^2)$}, and the leading coefficient
\begin{equation}
J(\omega) = \frac{4}{3} \int^1_0 dx\, \frac{\phi_\pi(x)}
                                              {1-\omega^2(2x-1)^2} \; ,
\label{J}
\end{equation}
samples the pion bound state through the distribution amplitude $\phi_\pi(x)$.
The normalization of the latter produces the model-independent result
\mbox{$J(0)=4/3$} for the case of equal virtuality.   The model-dependent
$J(1)$ for the asymmetric case has received the estimates: $2$ from use of
$\phi_\pi^{\rm asym}(x)$~\cite{Lepage:1980fj}, $1.8$ from a non-perturbative 
chiral quark model~\cite{Anikin:2000cx}, 1.6 from a fit to CLEO 
data~\cite{Anikin:2000cx}, and $4/3$ from analysis of the impulse 
approximation~\cite{Kekez:1999rw,dsegpglimit}.   A numerical
confirmation of the asymptotic behavior within the present DSE model 
requires a level of detail in the
representation of the analytic structure of dressed propagators and BS 
amplitudes that is not currently available.   However, for equal virtuality
there is no restriction, and our asymptotic result shown in 
Fig.~\ref{fig:piggff}~(right) produces \mbox{$J(0)=1.2$} which is within 10\%
of the $4/3$ result of Eq.~(\ref{J}) from the light-cone operator product
expansion.   
\section*{Acknowledgments}
I am  grateful to the University of Adelaide's Special Research Centre for 
the Subatomic Structure of Matter for their hospitality and financial 
support.  
This work draws heavily upon the results of collaborations with Pieter Maris
and Dennis Jarecke. I would also like to thank Craig Roberts, Tony Thomas, and
Tony Williams for useful discussions.  This work was partly funded by the 
National Science Foundation under grants Nos. PHY-0071361 and PHY-9722429, and 
benefited from the resources of the National Energy Research Scientific 
Computing Center and the Ohio Supercomputer Center.


\begin{thebibliography}{99}

\bibitem{Gockeler:1996wg}
M.~Gockeler, R.~Horsley, E.~M.~Ilgenfritz, H.~Perlt, P.~Rakow, G.~Schierholz and A.~Schiller,
Phys.\ Rev.\ D {\bf 53}, 2317 (1996);
G. Schierholz, these proceedings.

\bibitem{Weigel:1999pc}
H.~Weigel, E.~Ruiz Arriola and L.~Gamberg,
Nucl.\ Phys.\ B {\bf 560}, 383 (1999); 
E.~Ruiz Arriola, these proceedings.
%
\bibitem{Roberts:2000aa} 
C.~D.~Roberts and S.~M.~Schmidt, 
Prog.\ Part.\ Nucl.\ Phys. {\bf 45}, S1:1 (2000). 
%
\bibitem{Alkofer:2000wg}
R.~Alkofer and L.~von Smekal,
Phys. Rep. (2001), in press, [hep-ph/0007355].
%
\bibitem{Hecht:2001xa}
M.~B.~Hecht, C.~D.~Roberts and S.~M.~Schmidt,
Phys.\ Rev.\ C {\bf 63}, 025213 (2001).

\bibitem{MTpiK00}
P.~Maris and P.~C.~Tandy,
Phys.\ Rev.\ C {\bf 62}, 055204 (2000)
%
\bibitem{Ji:2001pj}
C.~Ji and P.~Maris,
in press, nucl-th/0102057.

\bibitem{Maris:1997tm}
P.~Maris and C.~D.~Roberts, Phys.\ Rev.\ C {\bf 56}, 3369 (1997).

\bibitem{Maris:1999nt}
P.~Maris and P.~C.~Tandy, Phys.\ Rev.\ C {\bf 60}, 055214 (1999).

\bibitem{Bender:1996bb}
A.~Bender, C.~D.~Roberts and L.~von Smekal, 
Phys.\ Lett.\ B {\bf 380}, 7 (1996).

\bibitem{Maris:2000bh}
P.~Maris and P.~C.~Tandy, Phys.\ Rev.\ C {\bf 61}, 045202 (2000).
%
\bibitem{MTinprogress} P. Maris and P.C. Tandy,
KSUCNR-109-00, in progress, proceedings of the Confinement Research Program at
the Erwin Schroedinger Institute for Mathematical Physics,
Vienna, May-July, 2000.
%
\bibitem{Maris:2000wz}
P.~Maris,
Nucl.\ Phys.\ Proc.\ Suppl.\  {\bf 90}, 127 (2000).
%
\bibitem{PDG}
Particle Data Group, C.~Caso {\it et al.},  
Eur.\ Phys.\ J.\ C {\bf 3}, 1 (1998).

\bibitem{Maris:1998hc}
P.~Maris and C.~D.~Roberts, Phys.\ Rev.\ C {\bf 58}, 3659 (1998).

\bibitem{Skullerud:2000un}
J.~I.~Skullerud and A.~G.~Williams, 
Phys. Rev. {\bf D 63}, 054508 (2001); 
J.~I.~Skullerud, D. B. Leinweber and A.~G.~Williams, 
hep-lat/0102013.
%
\bibitem{Maris:2000zf}
P.~Maris, 
nucl-th/0009064, to appear in conference proceedings;
%
C.~D.~Roberts,
nucl-th/0007054, to appear in conference proceedings.

\bibitem{Volmer} J.~Volmer, 
PhD thesis, Amsterdam 2000;
J.~Volmer {\it et al.}  [The Jefferson Lab F(pi) Collaboration],
Phys.\ Rev.\ Lett.\  {\bf 86}, 1713 (2001)

\bibitem{Farrar:1979aw}
G.~R.~Farrar and D.~R.~Jackson, Phys.\ Rev.\ Lett.\  {\bf 43}, 246 (1979).

\bibitem{JMT01prep} D. W. Jarecke, P. Maris and P. C. Tandy, 
KSUCNR-101-01, in progress, (2001).
%
\bibitem{cellocleo}
H.~J.~Behrend {\it et al.}  [CELLO Collaboration],
Z.\ Phys.\ C {\bf 49}, 401 (1991);
J.~Gronberg {\it et al.}  [CLEO Collaboration],
Phys.\ Rev.\  {\bf D57}, 33 (1998).

\bibitem{Lepage:1980fj}
G.~P.~Lepage and S.~J.~Brodsky,
Phys.\ Rev.\ D {\bf 22}, 2157 (1980).

\bibitem{Anikin:2000cx}
I.~V.~Anikin, A.~E.~Dorokhov and L.~Tomio, 
Phys.\ Lett.\ B {\bf 475}, 361 (2000).

\bibitem{Chase:1980ck}
M.~K.~Chase,
Nucl.\ Phys.\ B {\bf 167}, 125 (1980).

\bibitem{Kekez:1999rw}
D.~Kekez and D.~Klabucar, Phys.\ Lett.\ B {\bf 457}, 359 (1999).

\bibitem{dsegpglimit}
C.~D.~Roberts, Fizika B {\bf 8}, 285 (1999);
P.~C.~Tandy, Fizika B {\bf 8}, 295 (1999);
D.~Klabucar and D.~Kekez, Fizika B {\bf 8}, 303 (1999).

\end{thebibliography}
\end{document}